\documentclass{mpe_report}

\usepackage{psfig,graphicx,epsfig}
\usepackage{color}
\usepackage{amsmath,amssymb,epic,eepic,array}

\begin{document}

\pagenumbering{arabic}
\setcounter{page}{161}

 \renewcommand{\FirstPageOfPaper }{161}\renewcommand{\LastPageOfPaper }{164}\newcommand{\degrees} {^\circ}
\def\aap{A\&A}%
\def\apj{ApJ}%
\def\mnras{MNRAS}%

\title{The geometry of PSR B0031$-$07} 

\author{Roy Smits\inst{1} \and Dipanjan Mitra\inst{2} \and Ben
  Stappers\inst{3,4} \and Jan Kuijpers\inst{1} \and Patrick
  Weltevrede\inst{4}\and Axell Jessner\inst{5} \and Yashwant Gupta\inst{2}}

\institute{ Department of Astrophysics, Radboud University,
\phantom{$^1$} Nijmegen \and
National Center for Radio Astrophysics, Pune \and
Stichting ASTRON, Dwingeloo \and
Astronomical Institute ``Anton Pannekoek'', UvA \and
Max-Planck-Insitut f\"ur Radioastronomy, Bonn}
\maketitle

\begin{abstract}
  Here we present the results from an analysis of a
  multifrequency simultaneous observation of PSR~B0031$-$07. We have
  constructed a geometrical model, based on an empirical relationship
  between height and frequency of emission, that reproduces many of
  the observed characteristics. The model suggests very low emission
  altitudes for this pulsar of only a few kilometers above the star's
  surface.
\end{abstract}

\section{Introduction}
\begin{figure}
  {\rotatebox{0}{\resizebox{9cm}{!}{\includegraphics{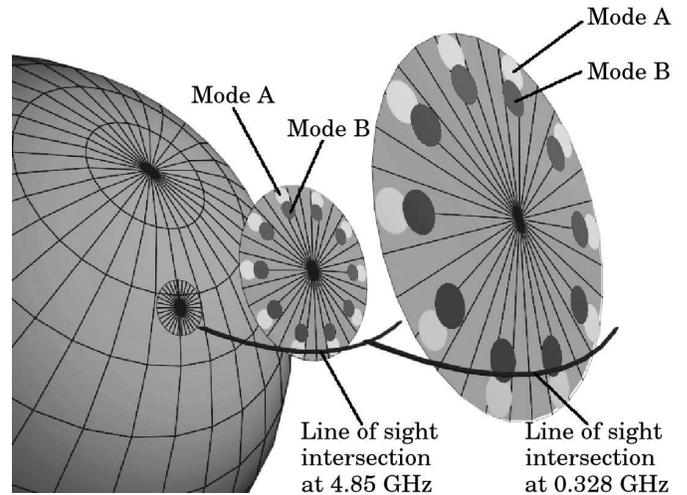}}}}
  \caption{Schematic overview of a geometrical model to
    explain the absence of one mode at high frequency. The two large discs are
    centered around the magnetic axis and represent the emission regions at
    two different frequencies, corresponding to two different altitudes above
    the pulsar surface. The smaller circles in the emission region represent
    the positions of the drifting sub-pulses, which rotate around the magnetic
    axis. The true number of sub-pulses is unknown. The different drift-modes
    are illustrated by different colours. Note that only one drift-mode is
    assumed to be active at a time.}
  \label{fig:pulsar}
\end{figure}

Pulsar B0031$-$07 is well known for its three modes of drifting sub-pulses.
They are called mode A, B and C and are characterised by their values for
$P_3$ of 12, 7 and 4 times the pulsar period, respectively (Huguenin et al.
1970).  This pulsar has been thoroughly studied at low observing frequencies
(Huguenin et al. 1970; Krishnamohan 1980; Wright 1981; Vivekanand 1995;
Vivekanand \& Joshi 1997, 1999; Joshi \& Vivekanand 2000), but only rarely at an
observing frequency above 1\,GHz (Wright \& Fowler 1981; Kuzmin et al.  1986;
Izvekova et al. 1993). Recently, Smits et al. (2005) analysed simultaneous
multifrequency observations from both the Westerbork Synthesis Radio Telescope
and the Effelsberg Radio Telescope and detected all three drift modes at
325\,MHz, but only detected drift mode A at 4.85\,GHz.  The pulses that were
classified as mode B or C at low frequency only showed non-drifting emission
at high frequency. On the basis of their findings, they suggested a
geometrical model where modes A and B at a given frequency are emitted in two
concentric rings around the magnetic axis with mode B being nested inside mode
A. As shown in Fig.~\ref{fig:pulsar}, this nested configuration is preserved
across frequency with the higher frequency arising closer to the stellar
surface compared to the lower one, consistent with the well known
radius-to-frequency mapping operating in pulsars.

Here we analyse new multifrequency observations of PSR~B0031$-$07, obtained
with the Giant Metrewave Radio Telescope, the Westerbork Synthesis Radio
Telescope and the Effelsberg Radio Telescope simultaneously. In total, the
observations contain 136\,000 pulses spread over 7 different frequencies. All
the observations were aligned by correlating long sequences of pulses with
pulses at an intermediate frequency that were obtained simultaneously.
Neglecting retardation and aberration, the accuracy of this alignment is
within 1\,ms. From these observations we attempt to restrict the geometry of
this pulsar and create a model that reproduces a great number of its observed
characteristics.

\section{Method}
To describe the observational drift of sub-pulses we use three
parameters, which are defined as follows: $P_3$ is the spacing, at the
same pulse phase, between drift bands in units of pulsar periods
($P_1$); this is the ``vertical'' spacing when the individual radio
profiles obtained during one stellar rotation are plotted one above
the other (stacked). $P_2$ is the interval between successive
sub-pulses within the same pulse, given in degrees, and $\Delta\phi$,
the sub-pulse phase drift, is the fraction of pulse period over which a sub-pulse
drifts, given in $\degrees$/$P_1$. Note that
$P_2=P_3\times\Delta\phi$. 

For both mode A and mode B, we calculated at each frequency the average
polarisation profiles of the total intensity, the profiles of the non-drifting
intensity, the values for $P_2$ and the \emph{fractional drift intensity},
which is a measure for how the line of sight intersection is away from the
center of the sub-beams. We then fitted the parameters of a simple geometrical
model, assuming constant emission heights based on an empirical relationship
between height and frequency of emission (Thorsett 1991) that reproduces the disappearing of
mode B at high frequencies by slightly moving the location of the emission
towards the magnetic axis.

\section{Results}
\begin{figure}
  {\rotatebox{0}{\resizebox{9cm}{!}{\includegraphics{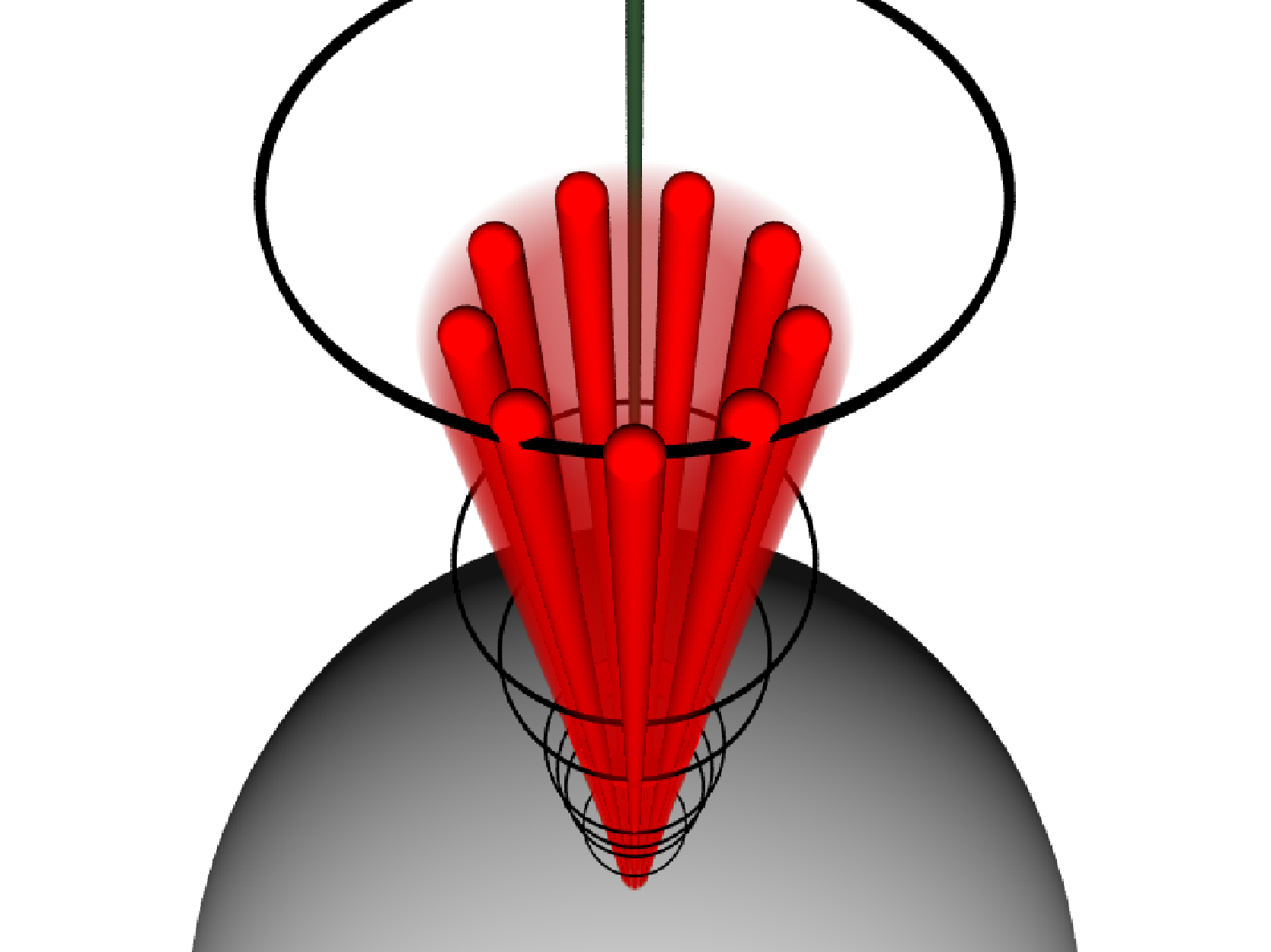}}}}
  {\rotatebox{0}{\resizebox{9cm}{!}{\includegraphics{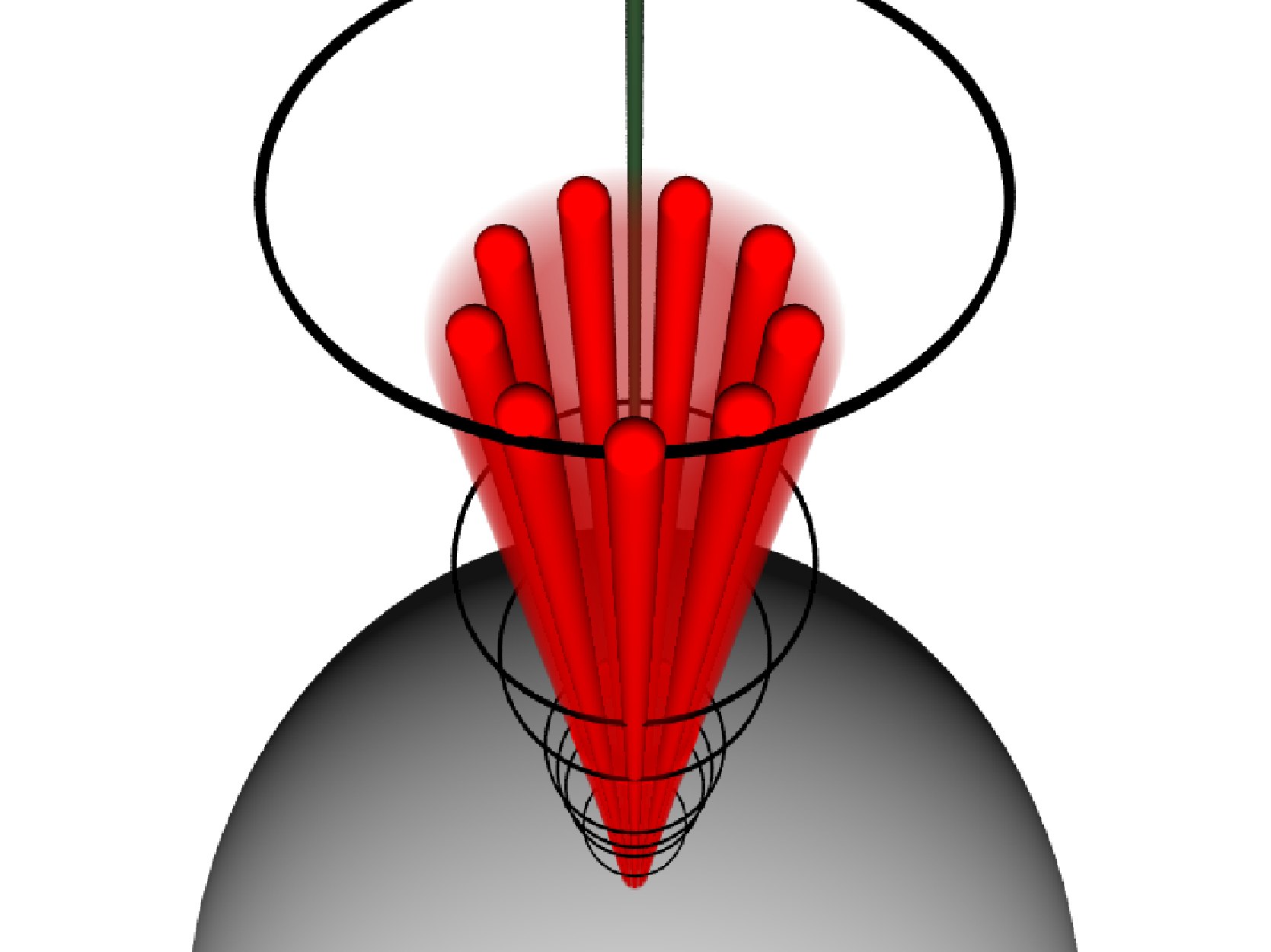}}}}
  \caption{Two close ups of the model of the emission zone of
  PSR~B0031$-$07. The vertical line is the rotation axis of the
  pulsar. The 7 circles indicate the line of sight trajectories
  corresponding to the 7 observed frequencies. The emission zone
  consists of 9 sub-beams surrounded by diffuse emission, shown as
  semi-transparent. Both images are to scale. The top image shows the
  location of the sub-beams during mode-A drift. The bottom image
  shows the location of the sub-beams during mode-B drift, which lie
  slightly closer to the magnetic axis.}
  \label{fig:model}
  \end{figure}

After optimizing the parameters of the model, it reproduces the observed
position angle sweep and the frequency dependence of the width of the average
intensity profile, the width of the average drift profile, $P_2$ and the
fractional drift intensity for drift modes A and B.
\begin{figure*}
  \begin{tabular}{cc}
      {\rotatebox{0}{\resizebox{7.5cm}{!}{\includegraphics{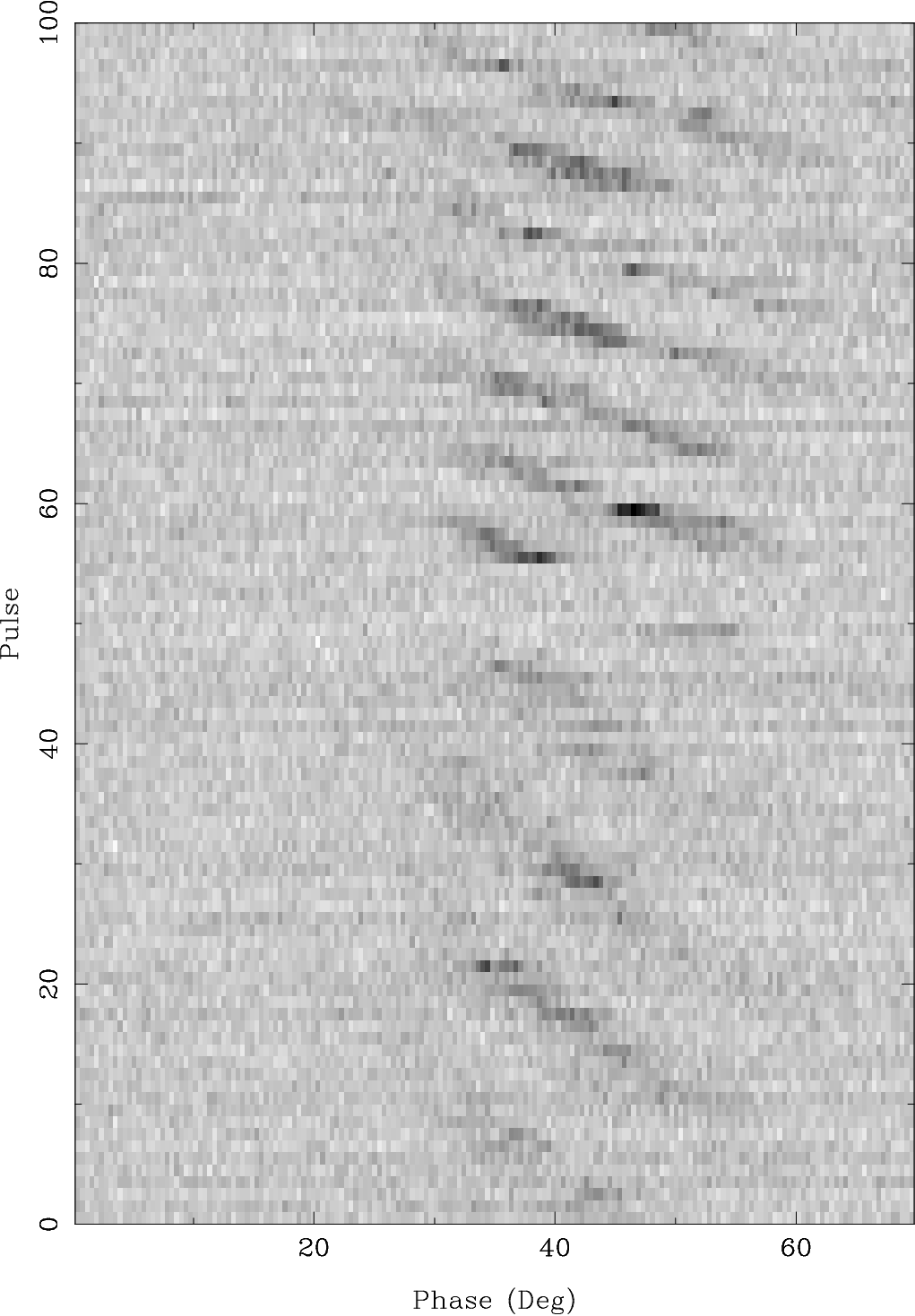}}}} &
      {\rotatebox{0}{\resizebox{7.5cm}{!}{\includegraphics{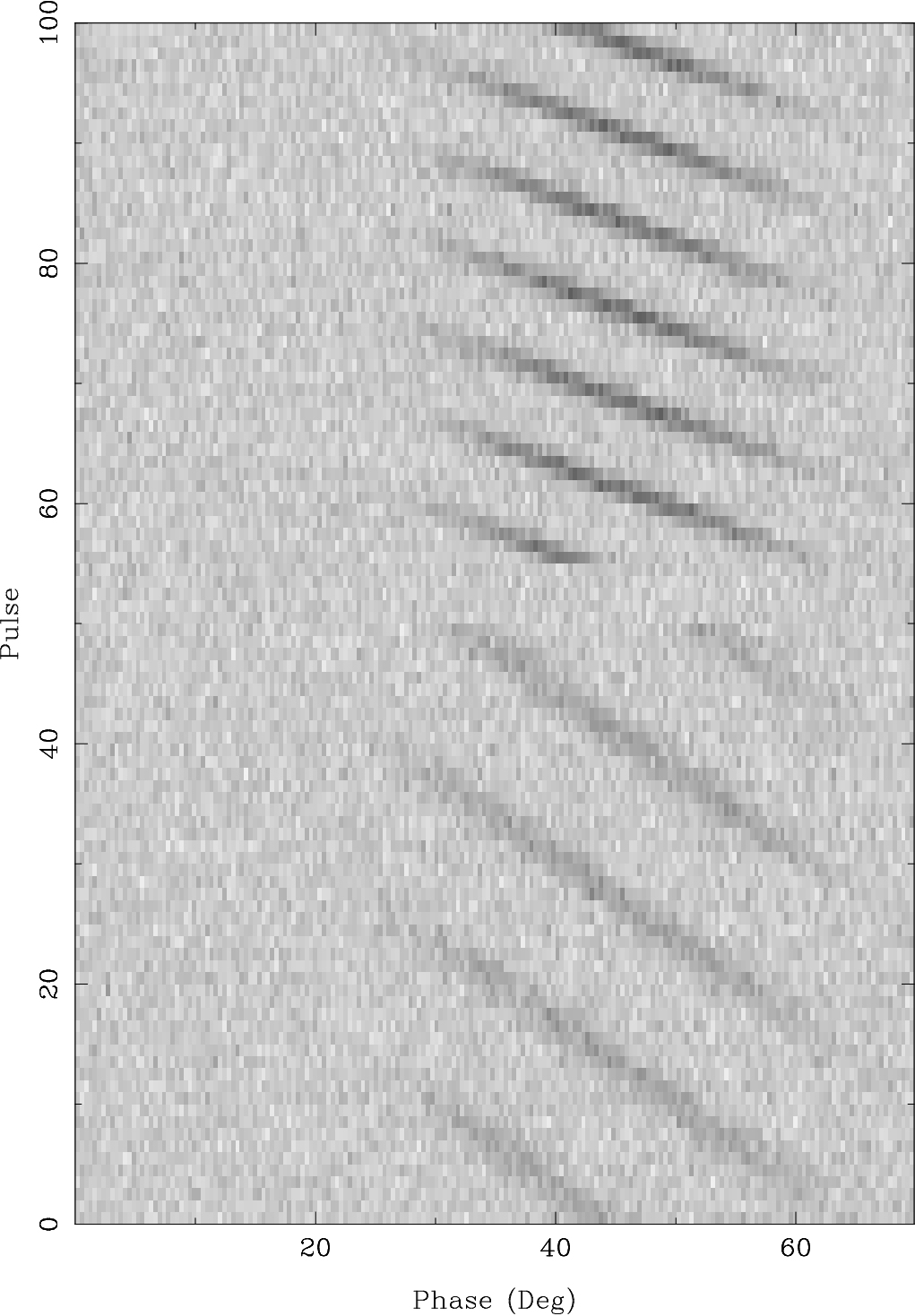}}}} \\
      {\rotatebox{0}{\resizebox{7.5cm}{!}{\includegraphics{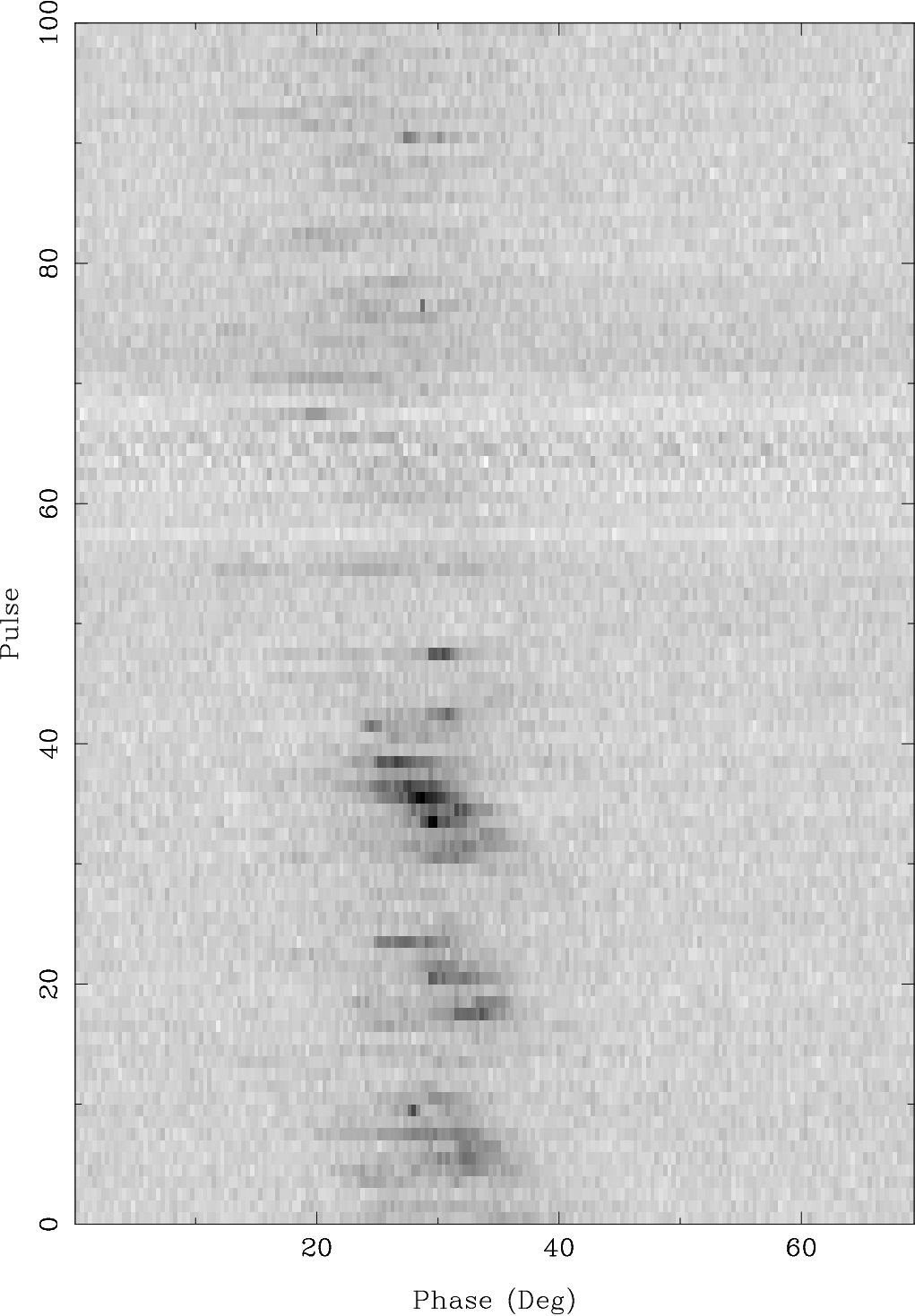}}}} &
      {\rotatebox{0}{\resizebox{7.5cm}{!}{\includegraphics{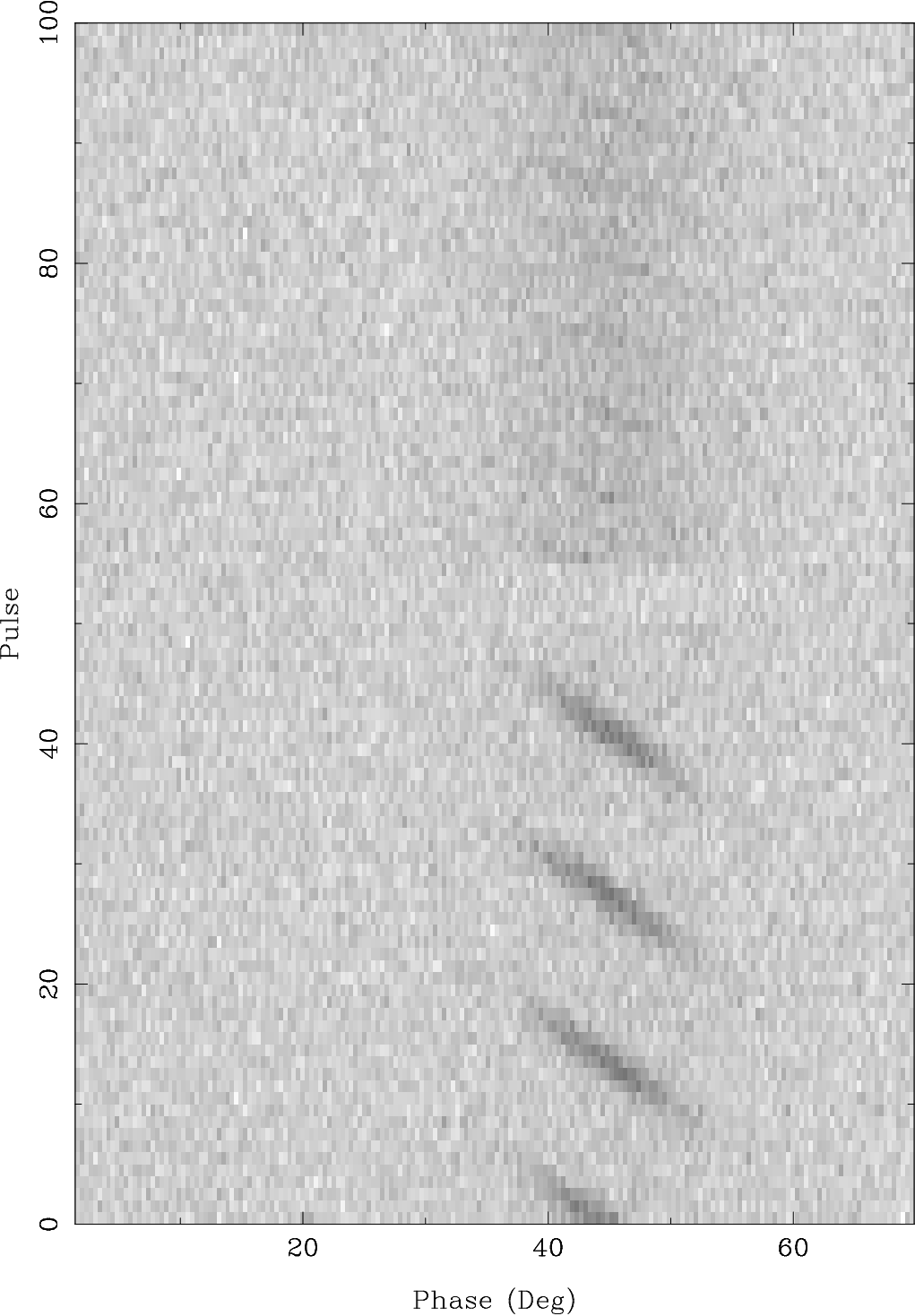}}}}
  \end{tabular}
  \caption{Gray scale plots of single pulses at two frequencies from
  the simultaneous observations (left panels) and from the model
  (right panels). The upper plots are at 243\,MHz and the bottom plots
  are at 4.85\,GHz. The first 50 pulses are in drift mode A, the following 5
  pulses are nulls and the remaining pulses are in drift mode B. To
  align the single pulses from the model with the single pulses from
  the observation, we assumed that the line of sight is closest to the
  magnetic axis at a pulse phase that corresponds to the center of the
  profile at 243\,MHz. This results in the offset between single
  pulses from the observation and from the model at 4.85\,GHz. There is
  some interference visible in the pulses 60 to 68 of the observation
  at 4.85\,GHz.}
  \label{fig:grayscales}
\end{figure*}
Fig.~\ref{fig:grayscales} shows the single pulses of the 243\,MHz observations
(top left), the 4\,850\,MHz observations (bottom left) with the corresponding
single pulses from the model to the right. The first 50 pulses are in mode A
and the last 50 pulses are in mode B. The model reproduces the disappearing of
the mode B drift at high frequency by changing the position of the emission
only slightly. The difference of the location of mode-A and mode-B emission is
illustrated in Fig.~\ref{fig:model}. The top image shows the true location of
the mode-A emission and the bottom image shows the true location of the mode-B
emission.

\section{Conclusions}
We can summarize the features of the geometrical model of
PSR~B0031$-$07 that is presented here as follows.
\begin{itemize}
\item The model reproduces the position angle sweep and the frequency
dependences of the width of the average intensity profiles, the width
of the average drift profiles, the fractional drift intensity and
$P_2$, for drift modes A and B of the single pulses of PSR~B0031$-$07.
\item The emission heights are very low. The high frequency emission
comes from a region a few kilometers above the surface of the
star. The low frequency emission comes from a region about 10
kilometers higher than the high frequency emission.
\item The parameters $\alpha$ and $\beta$ are approximately the same
and depending on the actual emission height, around 2$^\circ$ to
4$^\circ$.
\item The emission is centered around, or close to the last open field lines.
\item The emission from drift mode B comes from a region just slightly
closer to the magnetic axis than the emission from drift mode A.
\item Along with the drifting sub-pulses there is non-drifting emission
in the single pulses that becomes more significant towards higher frequencies.
\item Assuming that the observed drift speeds of the sub-pulses are not aliased, the number of sub-beams is around 9.
\end{itemize}
The model results in very low emission altitudes, ranging from 2.3 to
13.6\,km above the surface of the star. This is in strong contrast
with other emission heights that have been measured for pulsars, which
are typically some 10 to 1000\,km. However, by assuming a realistic
particle-density distribution near the polar cap, the emission heights
can become higher.

\begin{acknowledgements}
  The authors would like to thank J. Rankin, G. Wright and G. Melikidze for
  their rich suggestions and discussions. We also thank A. Karastergiou for
  his help on data alignment. This paper is based on observations with the
  100-m telescope of the MPIfR (Max-Planck-Institut f\"ur Radioastronomie) at
  Effelsberg, the Westerbork Synthesis Radio Telescope and the Giant Metrewave
  Radio Telescope and we would like to thank the technical staff and
  scientists who have been responsible for making these observations possible.
\end{acknowledgements}

              \clearpage

\end{document}